\documentclass[aps,prl,floatfix,reprint,superscriptaddress,showpacs]{revtex4-1}

\usepackage{amssymb}
\usepackage{graphicx}
\usepackage{amsmath}
\usepackage{color}

\newcommand{\sech}{{\rm sech}}
\begin{document}

\title{Kibble-Zurek scaling and its breakdown for spontaneous generation of Josephson\\ vortices in Bose-Einstein condensates}

\author{Shih-Wei Su}
\affiliation{Department of Physics, National Tsing Hua University, Hsinchu 30013 Taiwan}
\author{Shih-Chuan Gou}
\affiliation{Department of Physics and Graduate Institute of Photonics, National Changhua University of Education, Changhua 50058 Taiwan}
\author{Ashton Bradley}
\affiliation{Jack Dodd Centre for Quantum Technology, Department of Physics, University of Otago, Dunedin, New Zealand}
\author{Oleksandr Fialko}
\affiliation{Centre for Theoretical Chemistry and Physics, New Zealand Institute for Advanced Study,
Massey University (Albany Campus), Auckland, New Zealand}
\author{Joachim Brand}
\affiliation{Centre for Theoretical Chemistry and Physics, New Zealand Institute for Advanced Study,
Massey University (Albany Campus), Auckland, New Zealand}

\begin{abstract}
Atomic Bose-Einstein condensates confined to a dual-ring trap support Josephson vortices
as topologically stable defects in the relative phase. We propose a test of the scaling laws for defect formation
by quenching a Bose gas to degeneracy in this geometry. Stochastic Gross-Pitaevskii simulations reveal
a $-1/4$ power-law scaling of defect number with quench time for fast quenches, consistent with the Kibble-Zurek mechanism. Slow quenches show stronger quench-time dependence that is explained by the stability properties of Josephson vortices, revealing the boundary of the Kibble-Zurek regime. Interference of the two atomic fields enables clear  long-time measurement of stable defects, and a direct test of the Kibble-Zurek mechanism in Bose-Einstein condensation.
\end{abstract}
\date{Version of \today. }

\maketitle

A possible mechanism for the formation of domain structures in
the early universe was proposed by Kibble~\cite{Kibble}.
%jb It is
He argued that the Universe cooled down after the
hot Big Bang event and subsequently passed  through a
symmetry breaking phase transition
%with
at a critical temperature $T_{c}$.
Causally unconnected spatial domains settling into different vacua would lead to the formation of defects like domain walls, monopoles, strings, textures, etc \cite{Vilenkin94}.
Due to thermal fluctuations thwarting the emerging order, it was postulated that the number of defects eventually settled at the so-called Ginzburg temperature $T_{G} < T_c$.
\par
Later Zurek \cite{Zurek1} put forward an alternative argument focusing on the nonequilibrium aspect of the phase transition.
The density of the defects is determined at the critical temperature instead
and its number is scaled with the quench rate. The scaling exponent depends
on the critical exponents of the underlying phase transition. This scenario, known as the Kibble-Zurek mechanism (KZM) should equally apply to condensed matter phase transitions accessible to laboratory experiments \cite{Kibble07}. The KZM proved to be robust and was verified by a number of recent experiments on
annular Josephson tunnel junctions \cite{Monaco02,Monaco03,Monaco06,Monaco06prl} and
theoretical research on Bose-Einstein condensates (BEC) \cite{Zurek09,Das,Damski,delCampo}. It also extends to quantum phase
transitions~\cite{Dziarmaga08,Sabbatini11,Chen11}.
\par
%-------------------------------------------------------------------
\begin{figure}[tbp]
\includegraphics[width=.85\columnwidth]{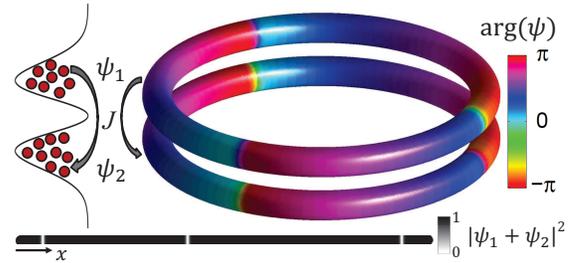}
\caption{(Color online) Schematic of the two linearly coupled BECs. The isosurface
shows the equilibrated condensate density profile
and the color shows a phase profile with three Josephson vortices resulting from a quench. The trapping potential is visualized on the
left. The interference pattern of the two atomic fields on the bottom shows clear evidence of the three Josephson vortices
located at the low density regions.}
\label{Fig.1}
\end{figure}
%-------------------------------------------------------------------
In the spirit of Kibble's argument one might expect the KZM to fail in the limit of slow quenches where the time scale of other processes occurring in the system dominates over the quench time.
Deviations from KZM predictions were observed in $^4$He experiments \cite{HeExpt} but the interpretation was controversial \cite{Rivers} and a manifestation of the Ginzburg temperature was ruled out in Ref.~\cite{Bettencourt}.
So far the transition between the regime of
KZ scaling and its breakdown has not been studied systematically.

In this Letter we investigate the robustness of the KZ scaling in a
system where departure from it can be understood in
%jb  details
detail because the
defects are easily quantified and are stable at the end of the quench.
This avoids the difficulty of counting the decaying population of defects \cite{Weiler08,Damski} or their remnants \cite{Sikivie82}.
To this end, we study two linearly coupled quasi-1D
atomic Bose gases in the ring configuration, as in Fig.~\ref{Fig.1}. A quench through the Bose-Einstein condensation phase transition
can generate Josephson vortices (JVs) confined between
two BECs \cite{Kaurov05,Brand1}. We show
%jb
that the number of JVs obeys the KZ
scaling law for fast quenches. On the contrary, for slow quenches, the
predicted behavior deviates substantially and we observe a much stronger quench-time dependence than expected for critical phenomena in our simulations. This is due to decay
processes occurring before the topological stability is established, in analogy to Kibble's arguments.

\par
The system under study can be realized by crossing a vertical Gaussian-Laguerre laser beam and two horizontal
sheet beams \cite{Campbell} to form an optical dipole trap or with rf-dressing on an atom chip \cite{Fernholz}.
Another way is trapping the atoms with two hyperfine states coupled via Raman transitions \cite{Sidorov}
in a single ring trap \cite{Campbell}. Along the $z$-axis the trapping potential can be treated as a double-well potential
as shown in Fig.~\ref{Fig.1}.
Assuming tight confinement, the transverse motion can be eliminated.
The resulting coupled Gross-Pitaevskii equations for the order parameter $\psi_1$ and $\psi_2$ in each ring assume the dimensionless form
\begin{eqnarray}
i\partial _{t}\psi _{j} &=&(\mathcal{L}_{j} -\mu)\psi _{j}-J\psi_{3-j}  \label{GPEd1}
\end{eqnarray}
where $\mathcal{L}_{j} = -\frac{1}{2}\partial _{xx}+g|\psi _{j}|^{2}$
($j=1,2$), $\mu$
%jb
is the chemical potential, and $J$
%jb is
the tunneling energy.
%jb The
Length, time, and energy are scaled by $a_h=\sqrt{\hbar/m\omega}$, $1/\omega $, and $\hbar\omega $, respectively, where $m$ is the atomic mass and $\omega $ is the transverse trapping frequency.
Accordingly, the dimensionless non-linear interaction strength $g$ is related to the s-wave scattering length $a$ by $g=2a/a_h$.
\par
Equation (\ref{GPEd1}) supports topological and non-topological defects in the form of
%jb
the JV and the dark soliton (DS), respectively,
$\tilde{\psi} _{1,2}=\sqrt{1+\nu}\tanh \left( p \tilde{x}\right) \pm iB\sech\left(
p \tilde{x}\right)$,
where $\nu=J/\mu$, and the scaling $x = \sqrt{\mu} \tilde{x }$ and ${\psi}_j= \sqrt{\mu/g} \tilde{\psi}_j$ has been applied.
Both the DS with $B=0$ and $p=\sqrt{1+\nu}$ and the JV with $p=2\sqrt{\nu}$ and $B=\sqrt{1-3\nu}$ for $\nu\le 1/3$ are
%jb
localised excitations
%jb
on the length scale $a_h(\sqrt{\mu}p)^{-1}$
above the vacuum where $\psi_1=\psi_2=\mathrm{const}$ \cite{Kaurov05}. The DS, where both components have identical profiles, is non-topological because it can continuously deform
to the vacuum by a family of moving ``grey'' solitons with decreasing energy \cite{PSbook}. Although they may be present transiently during quenches through the phase transition, DSs will thus not survive the final stage of cooling. Furthermore, for $\nu<1/3$, DSs are dynamically unstable with respect to decay into JVs, which have lower energy \cite{Qadir12}.
The stability properties of the JV, on the other hand, depend on the dimensionless parameter $\nu$ and may change during the quench. The JV bifurcates from the DS at $\nu=1/3$ as a time-reversal symmetry broken state (vortex and anti-vortex) with a characteristic phase winding of $2\pi$ around a point located between the two rings (see Fig.~\ref{Fig.1}), and only exists for smaller values of $\nu$.
From numerical simulations it is known that JVs can move with respect to the background BEC, although explicit solutions are unknown. For $1/5<\nu<1/3$
%jb perturbative
variational arguments indicate that the JV is energetically unstable \cite{Qadir12}. For $\nu<1/5$ where the JV resembles the Sine-Gordon soliton \cite{Kaurov05,Brand1}, the stationary solution is a metastable local energy minimum, since the energy increases with velocity. Thus, at sufficiently small $\nu$, JVs are topologically stable, enabling
experimental tests of the
KZ scaling by counting the number of JVs at the end of quench in a dual-ring BEC. The defects would be immediately evident by the interference images of two expanding atomic fields. The situation is strikingly different from a single
1D BEC where the
KZ scaling law was predicted to govern a transient population of
%jb
eventually decaying DSs,
%jb decaying finally,
which makes experimental detection more difficult \cite{Damski}.
\par
The nonequilibrium dynamics during the thermal quenches can be described by
the coupled stochastic Gross-Pitaevskii equations \cite{Ashton08,Blair08}:%
\begin{eqnarray}
d\psi _{j} &=& (i+\Gamma)\left[(\mu(t)-\mathcal{L}_j)\psi_j+J\psi_{3-j}\right]dt+dW_{j} ,  \label{SPGPE}
\end{eqnarray}%
where $\Gamma$ is the growth rate and $dW_{j}$ is the thermal noise satisfying the fluctuation-dissipation relation $\left\langle dW_{j}^{\ast }\left(x,t\right) dW_{k}\left( x^{{\prime }},t\right) \right\rangle =2\Gamma T\delta _{jk}\delta \left( x-x^{{\prime }}\right) dt$, with $T$ being the temperature in units of $\hbar \omega/k_{B}$. At the mean field equilibrium level the phase transition is described by the ground state of the energy
${\cal H}=\int dx\left[\frac{1}{2}|\partial_x\psi_1|^2+\frac{1}{2}|\partial_x\psi_2|^2 +V(\psi_1,\psi_2)\right]$,
where we seek the minimum of the potential
$V(\psi_1,\psi_2)\equiv\sum_{j=1,2}|\psi_j|^2\left[\frac{g}{2}|\psi_j|^2-\mu\right]-J\left[\psi_1^*\psi_2+\psi_2^*\psi_1\right]$ for $J>0$.
The symmetry $V(\psi_1,\psi_2)=V(\psi_2,\psi_1)$ imposes a common amplitude for the ground state fields. Taking $\psi_1=\sqrt{n}e^{i\phi_1}$,
$\psi_2=\sqrt{n}e^{i\phi_2}$, and $\Delta=\phi_1-\phi_2$, the minimum of $V=gn^2-2\mu n-2J n\cos{\Delta}$ occurs at $\Delta=0$, $n=(\mu+J)/g$, for $ \mu>-J$. At the critical point $\mu=-J$ the minimum is independent of $\Delta$ and each field breaks $U(1)$ symmetry.

The transition to the broken symmetry phase is simulated via Eq.~(\ref{SPGPE}) with time-dependent chemical potential
\begin{eqnarray}
\mu \left( t\right) &=&t/\tau _{Q},  \label{chemical potential}
\end{eqnarray}%
where $\tau _{Q}$ is the quench time. The quench starts from a thermal gas with a chemical potential $-\mu _{0}<0$, and ceases in the Bose-condensed phase at $\mu _{0}>0$. Due to
%jb
 inter-ring coupling, $\tilde{\mu}(t) = t/\tau_Q+J$ acts as the effective chemical potential; the precise location of the transition in a dynamical quench must be determined numerically.
We evaluate the total number of JVs during the quench with
$N_{JV}=\oint \left\vert d\left( \phi _{1}-\phi _{2}\right)
\right\vert /2\pi $.
The net number $N_{JV,net}=\left\vert\oint d\phi_1 - \oint d\phi_2\right\vert/2\pi$ is the difference between the number of clockwise and anti-clockwise vortices.

The KZ
theory applied to the BEC phase transition
gives the relaxation time and
healing length close to the critical point as
\begin{equation}
\tau =\tau _{0}|\tilde{\mu}|^{-1},\ \ \xi =\xi _{0} |\tilde{\mu}|^{-1/2},
\label{scaling1}
\end{equation}%
where $\xi _{0}$ and $\tau_{0}$ depend on the microscopic details of the system.
Following Eq.~(\ref{chemical potential}) and
the KZ
scenario \cite{zurek96}, we obtain the typical
size of the domains after the quench
\begin{eqnarray}
\hat{\xi} &=&\xi _{0}\left(\frac{\tau _{Q}}{\tau_{0}}\right)^{1/4},  \label{scaling2}
\end{eqnarray}%
where for our system $\tau _{0}=\Gamma ^{-1}$. When $\mu \left( t\right) $ exceeds $-J$, localized phase-domains start to grow in each ring. Typically a piece of the
(anti-)vortex will fall within a $\xi $-sized domain in which the phase is chosen randomly. Therefore, for small $J$, in a ring with circumference $C$, the total number of JVs is estimated to be
\begin{eqnarray}
\langle N_{JV}\rangle &\sim& C/\hat{\xi} = C\xi _{0}^{-1} \left( \frac{\tau
_{Q}}{\tau_{0}}\right)^{-1/4},  \label{totalscaling}
\end{eqnarray}%
and thus obeys the $-1/4$ power-law scaling with quench time.
The number of JVs thus shows a stronger quench-time dependence than the winding number of a single-ring BEC, which was predicted to scale with $\tau_Q^{-1/8}$ \cite{Das}.
\par
%-------------------------------------------------------------------
\begin{figure}[tbp]
\includegraphics[width=0.88\columnwidth]{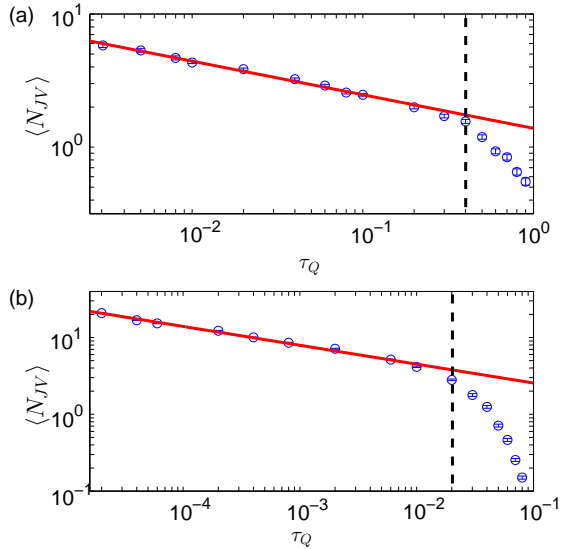}
\caption{(Color online) Scaling of the total number of JVs
with respect to $\protect\tau _{Q}$ at $J=5$ in (a)
and $J=25$ in (b) averaged over 500 trajectories of Eq.~(\ref{SPGPE}). The error bars indicate the standard deviations. The red
lines show the best power-law fit for fast quenches with exponents  $-0.2523\pm 0.0128$ in (a) and $-0.2456\pm 0.0131$ in (b), which agree
with the KZM prediction of -1/4. The dashed lines indicate the
critical quench time $\tau _{Q}^{crit}$ of Eq.~(\ref{condition}) for the breakdown of the %KZM
KZ scaling law.
}
\label{Fig.2}
\end{figure}
%-------------------------------------------------------------------
We consider a gas of $^{87}$Rb atoms with a transverse
confining frequency of $\omega=2\pi \times 200$Hz.
We numerically integrate Eq.~(\ref{SPGPE}) with $C=30$, $T=10^{-3}$,
and $g=0.05$, which are realistic
%jb
parameters with the set-up of Ref.~\cite{Henderson}.
The scaling in Eq.~(\ref{totalscaling}) is verified by averaging $
N_{JV}$ over 500 trajectories for $J=5$ and $25$.
The value of $\mu_0$ is chosen to be sufficiently large that the resulting defect number is independent of it.
As shown in Fig.~\ref{Fig.2}, the results
%jb of
for fast quenches
compare favorably with the KZM prediction, yet the number of JVs deviates from the KZ
scaling for slow quenches.
\par
The stability of a JV depends on conditions that change during the quench. According to the KZM, two different regimes exist: For early and late times during the quench, relaxation is efficient and fluctuations in the Bose gas follow the changing chemical potential adiabatically. However, when the diverging relaxation time $\tau$ of Eq.~(\ref{scaling1}) exceeds the time scale of the quench $\mu/\dot{\mu}$, fluctuations transiently freeze out and the system enters the \emph{impulse} regime. This occurs when
\begin{eqnarray}
\tau(\tilde{\mu}(\hat{t}))=|\tilde{\mu}/\dot{\tilde{\mu}}|_{t=\hat{t}}=\hat{t},
\end{eqnarray}
giving the freeze-out timescale $\hat{t}=\sqrt{\tau _{0}\tau _{Q}}$.
At the following impulse-adiabatic transition the frozen fluctuations are imprinted onto the forming BEC. We thus expect that the stability properties of defects formed at this transition point determine their survival during the adiabatic phase of the quench.
\par
%-------------------------------------------------------------------
\begin{figure}[tbp]
\includegraphics[width=.8\columnwidth]{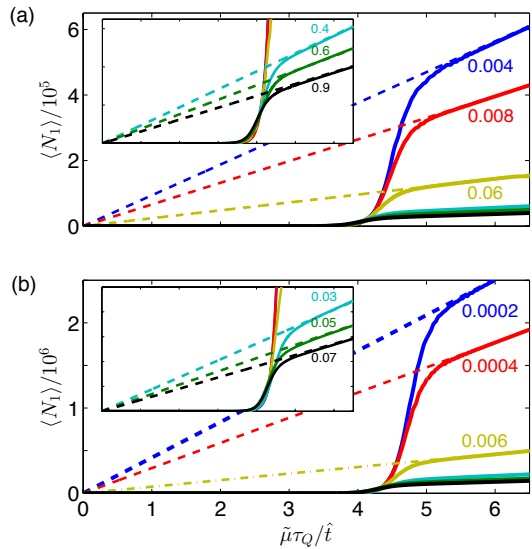}
\caption{(Color online) Particle number of component $\psi_1$ as a function of time
for $J=5$ in (a) and $J=25$ in (b). The vertical scale of the inset is magnified by a factor of 10 and reveal details for slow quenches. The color-coded labels show different $ \tau_Q$. Quenches with vastly different $\tau_Q$ show a knee structure characteristic of the impulse-adiabatic transition
with a rapid particle number increase  around
$\tilde\mu\tau _{Q} /\hat{t}=4.7$.
}
\label{Fig.3}
\end{figure}
\begin{figure}[t]
\includegraphics[width=1\columnwidth]{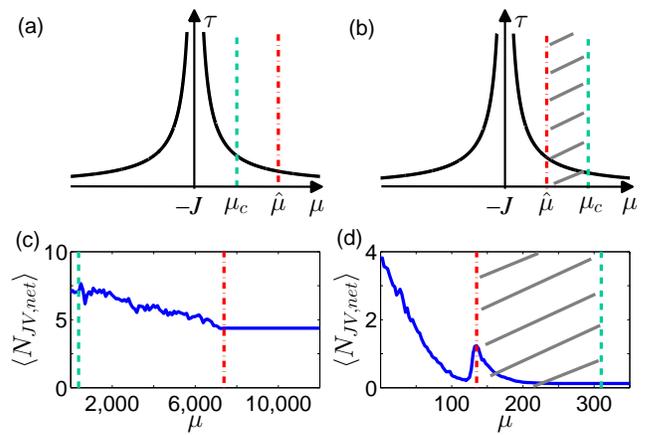}
\caption{(Color online)
Panels (a) and (b) show schematic plots of the relaxation time vs.\ chemical potential for fast and slow quenches, respectively. The KZ scaling is unaffected by the stability of JVs in (a), while the resulting number of
defects is affected by the decay happening in the shaded area in (b).
Panels (c), (d) show the net number of JVs for $\tau_Q=5\times10^{-5}$ (fast) and $0.08$ (slow) for $J=25$, respectively.
In (c) and (d), the locations of $\mu _{c}$ and $\hat{\mu}$ are obtained from $\nu_c=J/\mu_c$ and by reading the knee structure of the particle number [Fig.~3(b)], respectively.
%jb In (c),
For the fast quench in (c) the net number stabilizes right after $\hat{\mu}$, while for the slow quench it decays in the shaded region shown in (d).}
\label{Fig.4}
\end{figure}

%-------------------------------------------------------------------
In Fig.~\ref{Fig.3}, the impulse-adiabatic transition can be clearly observed from the particle number, $N_{1,2}(t)=\int \lvert\psi_{1,2}(x,t)\rvert^2dx$.
%jb The
A rapid increase of particle number
%jb is
takes place at $\tilde{\mu}= f \hat{t}/\tau_Q$, with $f \approx 4.7$. The value of $f$ appears
%jb
to depend weakly on the details
of the system, including the parameters $J$ and $\Gamma$, consistent with the theoretical argument that $\tilde{\mu}$ is relevant for the quench dynamics.
While the particle number is small in the impulse regime, it follows the dashed linear time-dependence in the adiabatic regime. Similar behavior was observed for a single ring BEC~\cite{Das}. Therefore we can predict the chemical potential at
the impulse-adiabatic transition as:
\begin{eqnarray}
\hat{\mu}&=&\mu \left( f\hat{t}-\tau_QJ\right) =f
\sqrt{\frac{\tau _{0}}{\tau _{Q}}}-J.  \label{chemical potential i-a}
\end{eqnarray}%
We denote the critical ratio of tunneling to chemical potential for JV stability by $\nu_c=J/\mu_c$, where $\mu_c$ is the stabilizing chemical potential for given $J$. As shown in Fig.~\ref{Fig.4}(a),
the defects are frozen in until $\hat{\mu}>\mu _{c}$ for fast quenches ensuring the topological protection of JVs and hence the KZM signature. However, for slow quenches the
impulse regime terminates earlier with $\hat{\mu}<\mu_c$, which causes the decay of the JVs
in the shaded region in Fig.~\ref{Fig.4} until the topological stability of JV is established at $\mu(t)=\mu_{c}$. Although the critical $\nu_c$ for a moving JV at finite temperatures
%jb remains unclear,
is unknown, we can estimate $\nu_c$ from the numerical simulations, at the point where
%jb the JV becomes unstable.
KZ scaling breaks down. From Eq.\ (\ref{chemical potential i-a}) we obtain the
criterion for obtaining stable JVs
\begin{eqnarray}
\tau _{Q}&<&\tau _{Q}^{crit}=\tau _{0}f^{2}(\nu_c/J)^{2}(1+\nu_c)^{-2}.
\label{condition}
\end{eqnarray}%
The value $\nu_c\simeq 0.0813 $ is obtained from the data for $J=5$, which
%jb gives
suggests $\tau_{Q}^{crit}=0.02$ for $J=25$,
as shown by the vertical dashed line plotted in Fig.~\ref{Fig.2}(b). This prediction agrees with the numerical data very
well. The critical quench time depends on the growth rate through $\tau _{0}$
and we have also verified the prediction of Eq.~(\ref{condition}) at different growth rates.
In the slow quench regime, as
%jb shown
seen in Fig.~\ref{Fig.2}, the defect number falls off more rapidly with quench time than expected from
%jb
the KZM. Since the variation is far from linear, we do not expect to enter a new regime of power-law scaling for slow quenches. Note that slow quenches show the same knee structure characterizing the impulse-adiabatic transition as the fast quenches that lead to
%jb the
KZ scaling (Fig.~\ref{Fig.3}).
We have also verified that $\langle N_{JV}\rangle$
%jb
continues to satisfy the KZ scaling
%jb
for slow quenches (solid line in Fig. 2), when counted immediately after the impulse-adiabatic transition at $\hat{\mu}$.
This supports our argument that the reduced defect number is due to thermal decay processes happening after the transition, and that
defect formation is unaffected by thermal fluctuations during freeze-out. Moreover, by varying the circumference, we verify that the KZM departure is not due to the finite-size effects discussed in Ref.~\cite{Weir,Biroli,zurek96}.
\par
%-------------------------------------------------------------------
%-------------------------------------------------------------------
%\par
For JVs, the slow quench regime is similar to Kibble's idea~\cite{Kibble}, where thermal fluctuations suffice to destroy the emerging order before the the system reaches the Ginzburg temperature. We observe a Ginzburg-like regime where thermal effects destroy the pattern of symmetry breaking inherited from criticality during the interval $\hat{\mu}<\mu<\mu_c$, shown in the shaded region of Fig.~\ref{Fig.4}(b). This scenario is consistent with the evolution of the net number of vortices during a quench shown in Fig.~\ref{Fig.4}(c) and (d). This measure is an indicator of the stability of individual JVs, unlike the total number that is affected by their pairwise annihilation (to which KZ scaling is immune).
\par
The absence of any clear cut evidence of cosmological nature and the difficulty
in observing the KZ scaling in condensed-matter experiments is usually not attributed to the failure of the mechanism
but may be explained by the decay of defects in the post-quench era \cite{Sikivie82,Weiler08,Damski}. This is circumvented
if the formed defects are topologically protected.
The defects observed in the successful experiments of Refs.~\cite{Monaco02,Monaco03,Monaco06,Monaco06prl} have this property, as do the JVs that are the subject of this Letter. While cosmological defects protected by topology may still survive in dark matter or dark energy, their
detection is difficult \cite{Pospelov}.
\par
Our work paves the way for a direct test of KZM in the Bose-Einstein condensation phase transition, by eliminating post-quench decay of defects \cite{Das}. The quench of $\mu$ could be supplanted by a controlled sweep of $J$, providing another knob for varying the quench time.
\par
Discussions with Brian Anderson and Ray Rivers are gratefully acknowledged. Simulations were performed on the Massey University CTCP computer cluster. SWS and SCG were supported by National Science Council, Taiwan (Grant No. 100-2112-M-018-001-MY3). OF and JB were supported by the Marsden Fund (Contract MAU0910). AB was supported by the Marsden Fund (Contract UOO162), and
a Rutherford Discovery Fellowship of the Royal Society of New Zealand (Contract UOO004).

\end{document}